\documentclass[useAMS,usenatbib]{mn2e}
\usepackage{graphicx}
\usepackage{times}

\def\farcs{\hbox{$.\!\!^{\prime\prime}$}}

\title[Spectroscopic orbits of 10 stars]{Spectroscopic orbits of ten nearby  solar-type dwarfs}

\author[Gorynya \& Tokovinin]{N.~A.~Gorynya$^{1,2}$\thanks{E-mail: gorynya@sai.msu.ru} 
 \& A.~Tokovinin$^{3}$\thanks{E-mail: atokovinin@ctio.noao.edu} \\
$^1$Institute of Astronomy of Russian Academy of Science, 48 Pyatnitskaya Str,
 109017 Moscow, Russia \\
$^2$ Lomonosov Moscow State University, Sternberg State Astronomical Institute,  Universitetskij prospekt, 13 Moscow
                119991, Russia \\
$^3$Cerro Tololo Inter-American Observatory, Casilla 603, La Serena, Chile\\
}

\begin{document}

\date{-}

\pagerange{\pageref{firstpage}--\pageref{lastpage}} \pubyear{2014}

\maketitle

\label{firstpage}

\begin{abstract}
Several nearby  solar-type dwarfs  with variable radial  velocity were
monitored to find their spectroscopic orbits.  Orbital elements of HIP
179, 1989,  2981, 5276, 6439,  11218, 21443, 96434 are  determined, as
well as tentative orbits for HIP  28678 and 41214.  We discuss each of
those objects.  Three of the four double-lined binaries are twins with
nearly equal components. All four  orbits with periods shorter than 10
days are circular, the remaining orbits are eccentric.
\end{abstract}


\section{Introduction}
\label{sec:intro}

Current  interest in  nearby  solar-like stars  is  largely driven  by
search of  exo-planets. These  stars are also  ideally suited  for the
study of binary statistics, which gives clues to the origin of stellar
and planetary  systems.  The 25-pc sample of  \citet{R10} was recently
extended to  a larger  volume to increase  the significance  of binary
statistics and to access the statistics of higher-order hierarchies \citep{FG67}.

Extensive data  on F- and  G-type stars within  67\,pc of the  Sun are
available  in the  literature  and  cover the  full  range of  orbital
periods.  The  detection of spectroscopic binaries is  mostly based on
the  Geneva-Copenhagen  Survey (GCS)  by  \citet{N04}. They  addressed
about 80\% of  the sample, typically with 2 or  3 radial velocity (RV)
measurements   per  star.    However,   many  spectroscopic   binaries
discovered by GCS from variable  RV or from the appearance of double lines
have no orbits determined so  far.  This leaves their periods and mass
ratios  indeterminate   and  adds  uncertainty   to  the  multiplicity
statistics.   A  large  number  of  these binaries  were  followed  by
D.~Latham at  Center for Astrophysics, leading to  hundreds of orbital
solutions (D.~Latham,  in preparation).  Still,  not all spectroscopic
binaries are covered.

Our aim is to complement  the existing work and to determine, whenever
possible,  spectroscopic orbits. We  focus on  stars with  the largest
(hence fastest)  RV variations and  derive here orbital  solutions for
some of them from the data of two seasons. The objects covered in this
study are listed in  Table~\ref{tab:list}, where visual magnitudes and
spectral types  are taken  from SIMBAD, trigonometric  parallaxes from
the {\it Hipparcos-2}  catalog \citep{HIP2},  and masses are  estimated from
  absolute  magnitudes  \citep{FG67a}.  All  stars  are bright;  their
  spectral types range from F2V to G5V.

\section{Observations}
\label{sec:obs}

The observations were conducted in  2012 and 2013 at the 1-m telescope
of  the Crimean  Astrophysical  Observatory sited  in Simeiz,  Crimea.
Radial   velocities  were   measured  by   the   CORAVEL-type  echelle
spectrometer, the Radial Velocity  Meter.  This instrument is based on
analogue  correlation of spectrum  with a  physical mask,  where slits
correspond to the spectral lines  \citep{RVM}. The RVs are measured by
fitting  Gaussian curves to  the observed  correlation dips,  with the
zero  point determined  from  observations of  RV standards.   Further
information on the  observing method and its limitations  can be found
e.g.  in \citep{TG01}.  The RV  precision reaches 0.3\,km~s$^{-1}$, but it is
worse  for  stars with  shallow  correlation  dips  and/or fast  axial
rotation.

\begin{figure*}
\centerline{
\includegraphics[width=17cm]{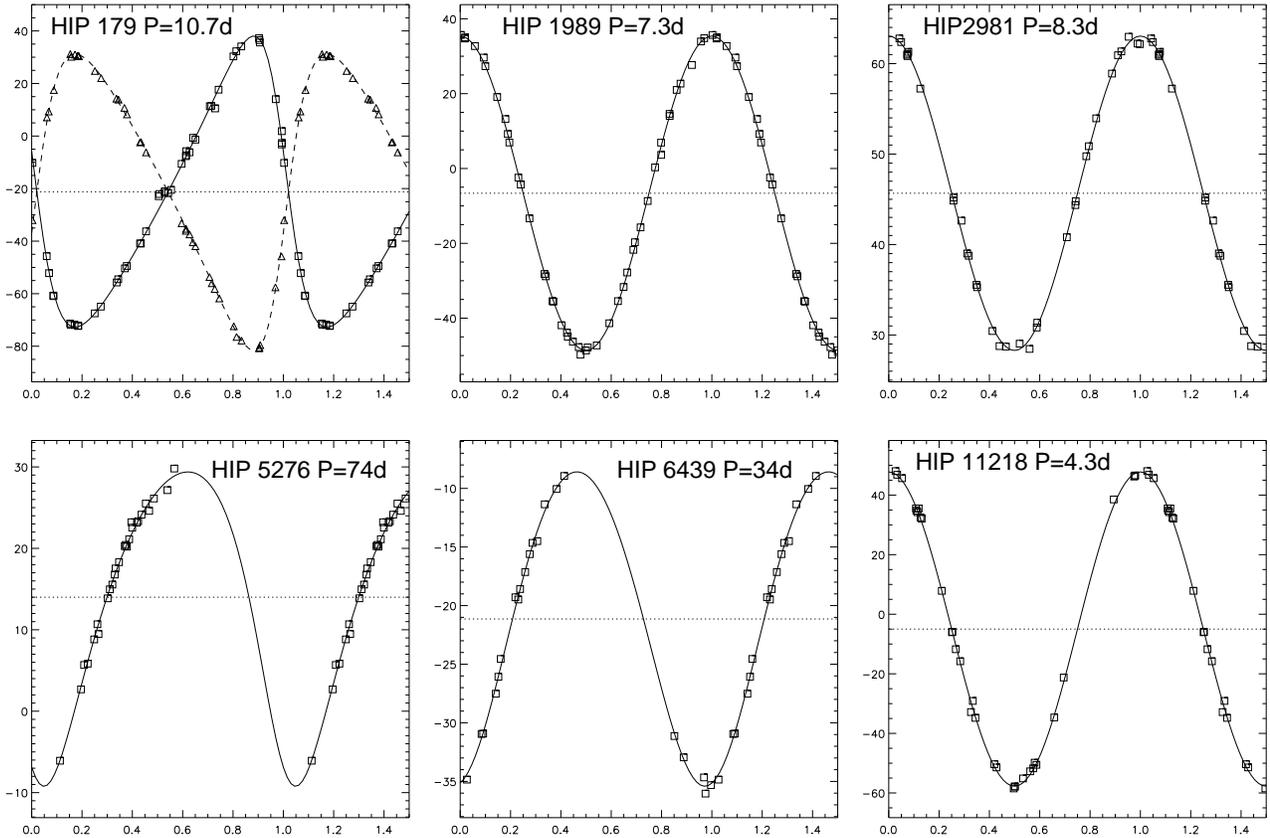}
}
\caption{Radial velocity curves. Each panel plots the orbital phase on
  the horizontal axis and the RV  (in km~s$^{-1}$) on the vertical axis.  The
  measurements are  plotted as squares  and triangles for  the primary
  and secondary  components, respectively.  The full  and dashed lines
  mark the  RV curves, the  horizontal dotted line corresponds  to the
  center-of-mass velocity.
\label{fig:orb1} }
\end{figure*}

\begin{figure*}
\centerline{
\includegraphics[width=12cm]{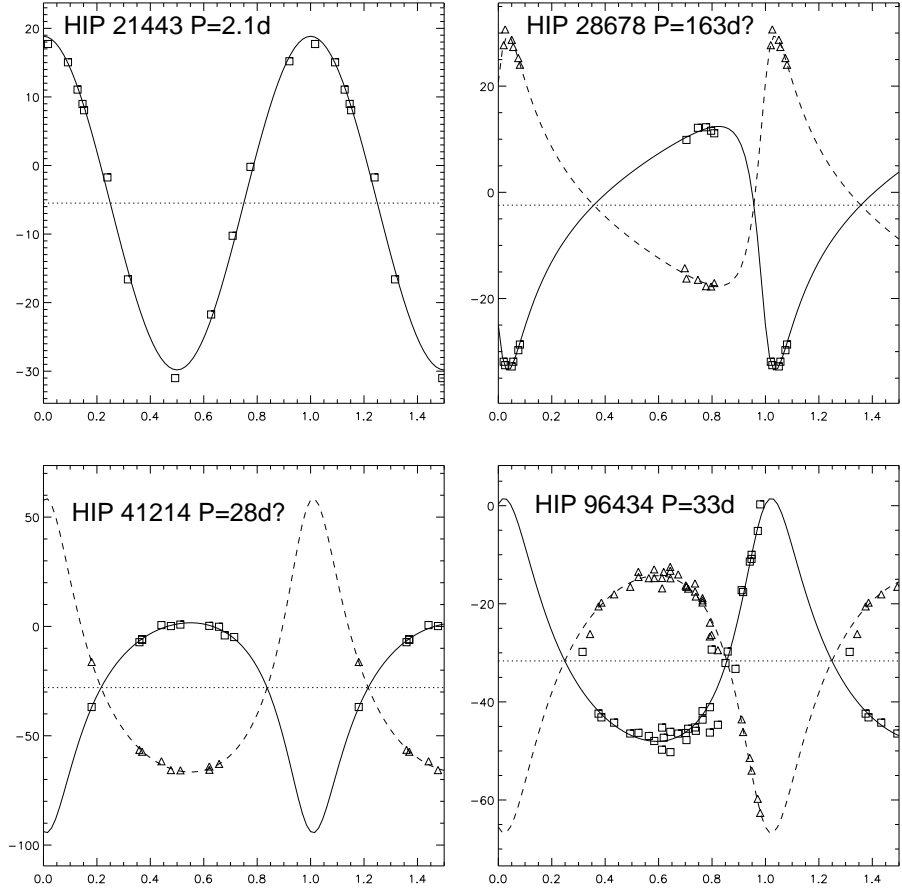}
}
\caption{Radial velocity curves (see Fig.~\ref{fig:orb1}). 
\label{fig:orb2} 
}
\end{figure*}

\begin{table}
\centering
\caption{List of stars with orbital solutions}
\label{tab:list}
\medskip
\begin{tabular}{rr c c c c } 
\hline
HIP & HD & $V$ & Spectral & $\pi_{\rm HIP}$ & $M_1$ \\
    &    & mag & type     & mas            & $M_\odot$ \\  
\hline
179 & 224974 & 6.90 & G0V  & 15.3 & 1.32 \\
1989 & 2085 &  7.89 & F5   & 16.8 & 1.22 \\
2981 & 3454 & 7.53  & F5   & 22.0 & 1.17 \\
5276 & 6611 & 7.24  & F5   & 19.4 & 1.29 \\
6439 & 8321 & 7.16  & F5   & 20.1 & 1.29 \\
11218 &14938& 7.20  & F5   & 18.0 & 1.35 \\
21443 & 28907 & 8.61& G5   & 15.5 & 1.11\\
28678& 41255& 7.47  & F9V  & 15.8 & 1.21 \\
41214 & 70937 & 6.03 & F2V & 15.8 & 1.61 \\
96434& 184962& 7.12 & F8   & 18.5 & 1.20 \\     
\hline
\end{tabular}
\end{table}

\section{New orbits}
\label{sec:orb}

Identifications and  basic parameters of stars  with orbital solutions
are  assembled in  Table~\ref{tab:list}.  Orbital  elements  and their
errors are  listed in  Table~\ref{tab:orb} in standard  notation.  Its
last  columns contain  the total  number of  RV measurements,  the rms
residuals to the orbit,  and mass estimates. For double-lined binaries
we  provide  $M \sin^3  i$, for  single-lined binaries  the minimum
secondary  mass  $M_{\rm  min}$  is  listed, computed  from  the  mass
function and the estimated primary  mass $M_1$ by solving the equation
$M_{\rm  min} =  4.695\,10^{-3}\, K_1  P^{1/3}(1 -  e^2)^{0.5}  (M_1 +
M_{\rm min})^{2/3}$. Individual measurements and their deviations from
the orbits  are listed in  Table~3, available in  full electronically.
Its columns contain  the heliocentric Julian day JD,  RV and its error
$\sigma$  in  km~s$^{-1}$,  residual  to  the orbit  O$-$C,  and,  for
double-lined  binaries,  the  component  identification ('a'  for  the
primary,  'b'  for the  secondary).   The  RV  curves are  plotted  in
Figs.~\ref{fig:orb1} and \ref{fig:orb2}.  In the orbital fits, RVs are
weighted as $1/(\sigma_i^2 +  0.3^2)$, where $\sigma_i$ are individual
RV errors  determined by  the Gaussian fits  of the  correlation dips,
with instrumental error of  0.3\,km~s$^{-1}$ added in quadrature.  RVs
derived from the unresolved  blended dips of double-lined binaries are
given a very  low weight by artificially increasing  the errors by few
tens  of km~s$^{-1}$.   Otherwise, the  double dips  are fitted  by two
Gaussian curves.  The formal errors are also  increased by
1  or  2  km~s$^{-1}$  for   stars  with  fast  rotation  and  shallow
correlation dips.  For circular orbits,  we fix the elements $e=0$ and
$\omega=0$ and fit the remaining four elements $P,T,K_1, \gamma$.

\begin{table*}
\centering
\caption{Orbital elements}
\label{tab:orb}
\medskip
\begin{tabular}{l ccc c ccc c c c }
\hline
HIP & $P$ & $T_0$ & $e$ & $\omega$ & $K_1$ & $K_2$ & $\gamma$ & $N$ & rms & Mass \\
    & days & +2400000 &  & deg     & km~s$^{-1}$ & km~s$^{-1}$ & km~s$^{-1}$      &     & km~s$^{-1}$ & $M_\odot$ \\   
\hline
179  & 10.6594  &  56233.373 & 0.343 & 78.1 & 55.39 & 56.40 & $-$21.23 & 46 & 1.00 & 0.64 \\
 & $\pm$0.0001  &      0.006 & 0.001 &  0.2 &  0.11 &  0.11 &     0.05 & 37 & 1.00 & 0.63 \\
1989 & 7.34780  & 56224.894  & 0     &  0   & 42.01 & -     & $-$6.61  & 44 & 0.81 & 0.56\\ 
 & $\pm$0.00003 &     0.002  & *     &  *   &  0.09 & -     &    0.06  &    &      & \\
2981 & 8.2559   & 56232.503  & 0     &  0   & 17.37 & -     &  45.67   & 33 & 0.49 & 0.20 \\  
  & $\pm$0.0004 &     0.013  & *     &  *   &  0.11 & -     &   0.08   &    &      &\\
5276 & 74.14    & 56282.15   & 0.231 & 151.4& 19.3  & -     & 14.0     & 27 & 0.55 & 0.56 \\  
 & $\pm$0.03    &     6.6    & 0.055 &  13.2&  3.8  &       &  4.4     &    &      &\\
6439 & 34.200   & 56261.5    & 0.065 & 191.5&  13.41& -     & $-$20.14 & 21 & 0.45 & 0.27 \\
 & $\pm$0.011   &     1.8    & 0.012 &  20.6&   0.17& -     &     0.33 &    &      &\\
11218 & 4.3202  & 56230.680  & 0     &  0   & 52.76 & -     & $-$4.98  & 33 & 1.04 & 0.64 \\
 & $\pm$0.0001  &     0.009  & *     &  *   &  0.15 & -     &    0.12  &    &      &\\
21443 & 2.0621  & 56604.522  & 0     &  0   & 24.33 & -     & $-$5.49  & 12 & 1.17 & 0.17 \\
 & $\pm$0.00013 &     0.004  & *     &  *   &  0.24 & -     &    0.18  &    &      &\\
28678& 163.2    & 56599.37   & 0.567 & 308.8& 23.01 & 23.59 & $-$2.42  & 11 & 0.48 & 0.48\\
  &  $\pm$3.0   &     0.45   & 0.050 &  4.4 &  0.38 &  0.39 &    0.11 & 12 & 0.63 &  0.47\\ 
41214& 27.901   & 56231.60   & 0.388 & 171.2& 48.0  & 62.6  & $-$28.0  & 11 & 0.86 & 1.73\\
  &  $\pm$0.028 &     0.71   & 0.067 &  6.8 &  5.6  &  7.0  &    1.1   &  9 & 0.73 & 1.32 \\ 
96434 & 33.542  & 56547.07   & 0.355 & 342.4& 24.78 & 26.24 & $-$31.65 & 35 & 1.27 & 0.19 \\  
 &  $\pm$0.043  &     0.15   & 0.010 &   2.1&  0.54 &  0.49 &     0.13 & 37 & 0.71 & 0.18 \\
\hline
\end{tabular}
\end{table*}

\begin{table}
\centering
\caption{Individual radial velocities and residuals (fragment)}
\label{tab:RV}
\medskip
\begin{tabular}{c  ccc c } 
\hline
JD  & RV  & $\sigma$ & O$-$C & Comp \\
+2400000 & km~s$^{-1}$ & km~s$^{-1}$ & km~s$^{-1}$  & \\
\hline
\multicolumn{5}{c}{HIP 179} \\
 56222.378 &$-$57.59  &   0.26  &   0.92 & b \\
 56222.392 &   14.02  &   0.30  &$-$0.59 & a \\
 56223.351 &$-$45.72  &   0.23  &   1.43 & a \\
 56223.360 &    7.08  &   0.25  &   1.45 & b \\
\hline
\end{tabular}
\end{table}



\section{Comments on individual objects}
\label{sec:stars}

We  discuss now  each star  individually.  The masses  of the  primary
components  were  derived  from  their  apparent  $V$  magnitudes  and
trigonometric parallaxes,  using standard relations  for main sequence
dwarfs   (corrections  were  made   for  magnitudes   of  double-lined
binaries). The existence of additional (tertiary) visual components is
noted.

{\bf HIP 179} has  a well-defined double-lined orbit with $P=10.65$\,d
and  nearly equal  RV  amplitudes, leading  to  the mass  ratio of  $q =
M_2/M_1  =0.982$ (GCS estimated $q=1.00 \pm 0.01$).  This binary is therefore a
{\em twin} with  identical components. The star is  elevated above the
main sequence by about 1.5\,mag, more than 0.75\,mag expected for a twin
binary.  Either  the components  are slightly evolved  (subgiants), or
the parallax is larger than measured by {\it Hipparcos}.

The semi-major  axis of the  spectroscopic orbit is about  2\,mas. The
star  was   observed  and  not  resolved   by  speckle  interferometry
\citep{Mason2001}  and by Robo-AO  \citep{RoboAO}. We  do not  see any
systematic RV  trend within a  year.  The existence of  any additional
tertiary components  thus appears unlikely, unless they  have low mass
and escaped direct resolution.

{\bf HIP 1989} has a circular orbit with $P=7.35$\,d with a relatively
massive secondary ($M_{\rm min} =  0.56\; M_\odot$).  The object is on
the main sequence.  This star is on the exo-planet program at the Keck
telescope   \citep{Isaacson10}   and    shows   the   RV   jitter   of
3.4\,m~s$^{-1}$ superposed  on the orbital RV variation.   The four RV
measurements  from   Keck  kindly  provided   by  D.~Fischer  (private
communication) match our orbit, but  are not included in the solution,
adding little to its  improvement.  To our knowledge, no spectroscopic
orbit  was published.   SIMBAD  lists the  RV of  $-47.4$\,km~s$^{-1}$
while  the   true  center-of-mass  velocity   is  $-6.6$\,km~s$^{-1}$.
Observations with  Robo-AO \citep{RoboAO} exclude  tertiary components
within the detection limit of that instrument.

{\bf HIP  2981} has a  8.25-day period. The circular  orbital solution
was imposed, as expected at such  short periods.  The object is on the
main sequence.  The  star is slightly metal-deficient, [Fe/H]=$-$0.58,
and has no additional known companions.

{\bf  HIP 5276}  has a  preliminary  orbit with  $P =  74$\,d. In  two
observing seasons, we  could not achieve a good  phase coverage, but
observed the ascending branch of  the RV curve three times, fixing the
period securely.   \citet{Nid02} published a single  RV measurement of
$-6.08$\,km~s$^{-1}$ made on JD~2451026;  curiously, they included the object
in  their   list  of   constant-velocity  stars.   We  used  this
measurement in  the orbital solution  presented in Fig.~\ref{fig:orb1}
(see the lowest data point).

The primary  mass of 1.29\,$M_\odot$ implies a  minimum secondary mass
of 0.56\,$M_\odot$. The system is  triple, it has a visual companion B
in the  2MASS catalog at 6\farcs17, 335\fdg4.   This companion was
confirmed as physical (co-moving) with Robo-AO \citep{RoboAO}.  With a
proper motion (PM)  of 131\,mas~yr$^{-1}$, the companion would  have moved if
it  were  a background  star.   The mass  of  the  visual companion  B
estimated from its photometry is about 0.3\,$M_\odot$.

{\bf HIP 6439} is similar to  the previous object in that the coverage
of the RV  curve is far from perfect, but  observations in two seasons
establish the  34-day period quite  well. The primary component  is on
the main sequence, the minimum  mass of the spectroscopic secondary is
0.27\,$M_\odot$.   The visual companion  at 71\arcsec  listed in  the WDS
\citep{WDS}  as STTA~16  is optical  because its  observed  motion is
opposite to the  PM of A.  The existence  of other tertiary components
was probed by speckle  interferometry at SOAR \citep{Hartkopf2012} and
with Robo-AO \citep{RoboAO}, none were found.

{\bf HIP  11218} has  a circular orbit  with $P=4.32$\,d. It  is about
1.1 mag above  the main sequence.  Fast  stellar rotation synchronized
with  the orbit  likely causes  the increased  chromospheric activity,
explaining why the object was detected  in X-rays by ROSAT.  It has no
additional  visual  companions,  as  far  as we  know.   SIMBAD  lists
metallicity [Fe/H]=$-$0.25.

{\bf  HIP 21443}  is  flagged in  the  GCS as  a single-lined  binary.
\citet{Guillout2009} confirm the fast  RV variability and relate it to
the  detection of  this  star  by ROSAT  (their  two RV measurements  are
included in the present orbit with an offset of +7\,km~s$^{-1}$). The lithium
line and  H$\alpha$ emission indicate  that HIP~21443 is  young.  Fast
rotation matches  the 2\,d  period found here,  although the  orbit is
based on only 12 RV  measurements.  

The 2MASS  catalog contains a  star at 5\farcs3, 12\fdg6  from the
main  target.  This  companion  was measured  with  Robo-AO at  similar
position. However, the  PM of HIP~21443 is small,  20\,mas~yr${-1}$, and the
field is  rather crowded.   It remains to  be established  whether the
2MASS companion is bound or is just a background star.

{\bf HIP 28678} shows double lines.  In two seasons, we covered only a
small fraction of  its RV curve, so the  present orbit with $P=163$\,d
is rather tentative.   The mass ratio $q=0.98$ found  here is close to
$q=0.947$  estimated in  the GCS  from 11  observations.   The orbital
period implies  a semi-major axis of  12\,mas, so the  binary could be
marginally resolvable by speckle interferometry.  It was observed with
speckle  interferometry at  the  4.1-m SOAR  telescope  in 2010.9  and
2014.05 and  found unresolved  \citep{Hartkopf2012}.  In the  light of
the present orbit, we re-examined  the speckle data and found a slight
asymmetry indicative  of marginal resolution,  although no measurement
of the separation  could be made.  The spectral type  is F9V, the star
is moderately elevated above the  main sequence owing to its binarity.
No additional visual companions are listed in the WDS.

{\bf HIP  41214} (HR~395) is a  double-lined binary.  The  data do not
yet  constrain  the orbit  well.   The  28-day  period suggested  here
represents the  observations, but  may be not  a unique  solution. The
orbit  implies  $q=0.77$,  while   GCS  estimated  $q=0.86$  from  two
observations.  The  star appears slightly evolved. It  is monitored at
Keck for exo-planets \citep{Isaacson10}. Note that the primary mass of
1.73\,$M_\odot$  slightly exceeds  1.61\,$M_\odot$ estimated  from the
luminosity (a magnitude difference of 0.85\,mag was adopted to subtract
the  light of  the secondary).  This means  high  orbital inclination.
This  star  is not  identified  as a  variable  in  the ASAS  database
\citep{ASAS}.   Although  the  ASAS   photometry  does hint on
occasional minima, they do not follow the 27.9-d periodicity.


{\bf HIP 96434.} In two seasons, we were able to derive a double-lined
orbit with a 30-d period  and moderate eccentricity. The RV errors are
large because  the correlation dips  are shallow.  The  components are
nearly equal  in mass, $q=0.96$.  The GCS  recognized the double-lined
nature  of this  binary  but did  not  estimate its  mass ratio.   The
semi-major axis of 4.5\,mas puts this pair beyond the reach of speckle
interferometry, even  at 8-m telescopes.  The WDS  companion H~5104 at
39\arcsec, 138\degr  ~is optical, as  evidenced by its  relative motion
which  simply reflects  the  PM  of the  main  target.  No  additional
components were found with Robo-AO \citep{RoboAO}.

\section{Conclusions}
\label{sec:concl}

The  database  on  solar-type  binaries  within  67\,pc  \citep{FG67a}
suffers  from  the missing  data  (periods  and  mass ratios)  on  260
spectroscopic binaries  discovered by the  GCS. Our work  reduced this
number by  4\%; its  results are already  included in  the statistical
analysis  of  this  sample  \citep{FG67}.  Several  other  stars  with
variable RV  and/or double  lines were observed,  but do not  have yet
sufficient  data to  derive  their  orbits. We  plan  to continue  the
observations.

\section*{Acknowledgments}

This work  was supported by  the Russian Foundation for  Basic Research
(project code 11-02-00608). We thank  the administration of the Simeiz
Section  of  the  Crimean  Astrophysical  Observatory  for  allocating
observing time on the 1-m telescope.




\tiny{

\begin{verbatim}

Table 3 (full version}
----------------------------------------------
HJD            RV         Err       O-C  Comp
----------------------------------------------

HIP 179

 56222.378   -57.590     0.260     0.920  b
 56222.392    14.020     0.300    -0.588  a
 56223.351   -45.720     0.230     1.434  a
 56223.360     7.080     0.250     1.449  b
 56224.341   -71.350     0.250     0.850  a
 56224.350    31.190     0.460     0.481  b
 56226.305    14.210     0.270     0.190  b
 56226.315   -55.740     0.430    -0.029  a
 56227.339   -40.890     0.210    -0.712  a
 56227.339    -2.250     0.260    -0.310  b
 56228.347   -21.610    30.200     2.013  a
 56230.322    11.480     0.280    -0.594  a
 56230.343   -56.030     0.350    -0.482  b
 56232.341    37.220     0.390     0.140  a
 56232.357   -80.510     0.270    -0.068  b
 56233.304    -3.200    21.550    -2.037  a
 56233.311    -2.630    21.460    -0.992  a
 56235.327    30.570     0.260    -0.477  b
 56235.343   -72.290     0.200     0.248  a
 56237.301    10.710     0.470     1.695  b
 56237.322   -50.370     0.260     0.254  a
 56239.277   -20.510    30.220    -1.519  a
 56240.309    -1.380     0.250    -0.841  a
 56240.309   -41.930     0.240     0.371  b
 56241.290    17.630     0.230    -0.272  a
 56241.321   -61.810     0.340    -0.141  b
 56242.251    34.100     0.290    -0.029  a
 56242.263   -77.860     0.230    -0.104  b
 56526.557   -22.910    90.240     4.974  a
 56526.585   -22.080    90.300     5.336  a
 56527.527   -10.590     0.370     0.575  a
 56527.527   -33.190     0.430    -1.709  b
 56532.558     9.340     0.300    -0.033  b
 56532.573   -52.190     0.250    -0.211  a
 56533.493   -71.870     0.300     0.467  a
 56533.508    30.150     0.370    -0.716  b
 56534.506   -67.490     0.300     0.071  a
 56534.515    24.760     0.350    -1.084  b
 56535.501   -54.510     0.370     0.290  a
 56535.512    13.850     0.400     1.064  b
 56536.426   -40.870     0.300    -0.129  a
 56536.426    -2.400     0.360    -1.033  b
 56537.475   -21.040     2.200     2.490  a
 56538.524    -6.190     0.500    -1.101  a
 56538.524   -37.430     0.380     0.238  b
 56539.590   -58.200     0.720    -0.254  b
 56539.606    10.500     5.990    -4.630  a
 56540.502    32.260     0.320     1.161  a
 56540.515   -76.470     0.400    -1.752  b
 56541.500    35.600     0.400    -1.098  a
 56541.515   -79.580     0.310     0.455  b
 56542.527   -10.230     0.410    -2.126  a
 56542.527   -32.000     0.480     2.598  b
 56544.467   -72.190     0.320     0.345  a
 56544.479    30.600     0.360    -0.372  b
 56544.488    30.360     0.270    -0.586  b
 56545.423   -64.930     0.250    -0.272  a
 56545.438    22.030     0.230    -0.771  b
 56546.524   -49.460     0.330    -0.030  a
 56546.532     8.240     0.290     0.883  b
 56602.325    -7.660     0.530     0.242  a
 56602.325   -36.070     1.250    -1.267  b
 56602.332    -7.330     0.310     0.446  a
 56602.332   -35.340     0.270    -0.408  b
 56602.340    -5.750     0.510     1.882  a
 56603.323   -53.650     0.490    -0.008  b
 56603.337    11.320     0.240     0.455  a
 56604.338    30.310     0.410     1.135  a
 56604.354   -72.400     0.480     0.422  b
 56605.424   -80.780     0.240    -0.217  b
 56605.438    36.480     0.400    -0.419  a
 56606.372   -45.790     0.270    -3.373  b
 56606.386     1.850     0.260     3.223  a
 56607.369   -60.720     0.480    -1.488  a
 56607.375   -60.920     0.250    -1.479  a
 56607.390    17.500     0.360    -0.696  b
 56608.271   -71.900     0.300     0.806  a
 56608.284    31.020     0.250    -0.163  b
 56611.297   -36.280     0.250     0.384  a
 56611.297    -6.210     0.350    -0.691  b
 56612.234   -21.720    10.180    -0.632  a
 56613.279   -40.450     0.270    -0.191  b
 56613.288    -0.680     0.280     1.699  a

HIP 1989

 56222.422   -27.800     0.270     0.527  
 56223.411     6.910     0.190     1.011  
 56226.330     6.940     0.300    -0.575  
 56227.355   -28.200     0.290    -0.226  
 56228.406   -49.760     0.320    -1.543  
 56230.368    -8.690     0.300    -0.745  
 56232.378    34.730     0.220    -0.378  
 56232.383    35.040     0.210    -0.047  
 56235.383   -44.960     0.230    -0.627  
 56237.343   -19.710     0.280     1.327  
 56239.367    34.830     0.190     0.199  
 56240.324    27.330     0.230    -0.052  
 56241.357    -4.300     0.300    -0.191  
 56242.284   -35.450     0.260    -0.729  
 56527.544     9.220     0.220     0.164  
 56532.587    22.680     0.310    -0.499  
 56533.520    35.680     0.300     0.293  
 56534.582    19.110     0.400     0.397  
 56535.524   -13.350     0.410    -0.114  
 56536.461   -41.910     0.330    -0.896  
 56537.488   -47.300     0.320    -0.168  
 56538.558   -21.740     0.380     0.800  
 56539.615    14.600     0.470     0.525  
 56539.618    14.030     0.310    -0.138  
 56540.530    33.950     0.310     0.118  
 56541.537    29.630     0.230     1.286  
 56542.548    -2.440     0.300    -0.795  
 56544.522   -48.640     0.230    -0.020  
 56545.455   -35.450     0.230     0.537  
 56546.540     0.300     0.300     0.494  
 56602.354   -35.590     0.180    -0.138  
 56603.356   -47.820     0.190     0.761  
 56604.403   -31.670     0.260    -0.215  
 56605.500     3.640     0.290    -2.372  
 56606.396    27.660     0.260    -2.603  
 56607.409    32.710     0.200     0.114  
 56608.300    13.210     0.250     1.857  
 56611.334   -41.360     0.260     0.353  
 56612.245   -15.740     0.300    -0.362  
 56613.300    21.000     0.340     0.812  

HIP 2981

 56222.473    49.760     0.180     0.287  
 56226.371    44.850     0.300    -0.028  
 56226.379    45.190     0.290     0.418  
 56230.381    44.790     0.260    -0.114  
 56232.410    62.220     0.240    -0.781  
 56235.402    35.280     0.280    -0.077  
 56237.358    30.820     0.340    -0.069  
 56241.374    61.090     0.200    -0.086  
 56241.374    60.830     0.260    -0.346  
 56241.402    61.300     0.290     0.295  
 56527.584    44.360     0.260    -0.411  
 56532.595    35.540     0.350    -0.032  
 56533.560    28.720     0.290     0.019  
 56534.589    31.360     0.400     0.345  
 56535.559    40.810     0.550    -0.300  
 56536.522    53.950     0.320     0.453  
 56537.588    62.970     0.360     0.662  
 56538.577    60.970     0.750    -0.265  
 56540.546    39.010     0.300    -0.085  
 56541.606    28.770     0.380    -0.740  
 56542.592    28.460     0.340    -1.040  
 56544.535    50.870     0.300     0.358  
 56545.492    60.930     0.320     0.540  
 56546.578    62.780     0.340     0.348  
 56598.389    38.760     0.230     0.291  
 56603.394    61.350     0.240     0.236  
 56604.428    62.380     0.310     0.169  
 56606.404    42.640     0.300     1.168  
 56607.422    30.440     0.220    -0.437  
 56608.315    29.040     0.240     0.602  
 56611.344    58.900     0.470     0.039  
 56612.254    62.150     0.250    -0.892  
 56613.311    57.230     0.320    -0.691  

HIP 5276

 51026.500    -6.080     0.100    -0.044  
 56222.543     2.680     0.270    -0.366  
 56223.454     5.680     0.250     1.219  
 56224.522     5.840     0.290    -0.256  
 56226.420     8.820     0.180    -0.087  
 56227.372    10.680     0.230     0.420  
 56230.398    13.860     0.280    -0.409  
 56232.480    16.780     0.210     0.028  
 56235.427    20.270     0.250     0.393  
 56237.370    23.210     1.580     1.518  
 56239.387    23.340     0.210    -0.035  
 56240.367    24.140     0.190     0.021  
 56527.598    14.970     0.260    -0.082  
 56532.604    20.210     0.320    -0.262  
 56535.570    23.160     0.570     0.086  
 56540.577    26.120     0.320    -0.353  
 56544.549    27.150     0.320    -1.132  
 56546.585    29.800     0.280     0.911  
 56598.401     9.480     0.260    -1.233  
 56602.492    15.550     0.180    -0.402  
 56603.409    17.530     0.280     0.524  
 56604.442    18.300     0.320     0.158  
 56606.415    20.420     0.200     0.265  
 56607.436    21.120     0.250     0.002  
 56608.326    22.540     0.260     0.626  
 56612.328    25.500     0.290     0.495  
 56613.322    24.600     0.370    -1.050  

HIP 6439

 56223.466   -32.950     0.210     0.169  
 56226.431   -36.040     0.270    -0.629  
 56230.418   -30.920     0.240    -0.063  
 56232.491   -26.070     0.300    -0.045  
 56235.438   -18.590     0.210     0.016  
 56240.387   -10.050     0.210    -0.106  
 56241.434    -8.940     0.300     0.172  
 56540.607   -24.550     0.290     0.690  
 56542.601   -19.290     0.400     0.880  
 56544.560   -15.610     0.300     0.034  
 56545.584   -14.500     0.220    -0.870  
 56546.594   -11.360     0.280     0.582  
 56598.412   -31.130     0.260    -0.233  
 56602.435   -34.660     0.200     0.750  
 56603.384   -35.320     0.230    -0.111  
 56604.415   -34.850     0.320    -0.450  
 56606.408   -30.950     0.230     0.347  
 56608.337   -27.520     0.200    -0.593  
 56611.401   -19.480     0.300    -0.287  
 56612.338   -17.140     0.300    -0.139  
 56613.332   -14.650     0.260     0.233  

HIP 11218

 56223.483   -29.110     1.340     2.493  
 56224.544   -49.830     1.340     1.426  
 56226.477    48.050     1.320     1.038  
 56227.446    -6.050     1.350    -0.578  
 56232.530   -51.430     1.370     1.044  
 56235.493    34.270     1.370    -0.521  
 56237.429   -52.730     1.340     1.033  
 56240.412    -5.910     1.390    -0.036  
 56241.466   -58.480     1.350    -0.748  
 56242.324   -21.250     1.370     1.524  
 56533.573    34.760     1.370    -0.897  
 56544.580   -34.660     1.440    -0.804  
 56545.601    38.530     1.390     1.993  
 56546.604    32.470     1.350     0.456  
 56598.369    35.520     0.360    -0.515  
 56598.467    32.100     0.310     1.194  
 56602.367    46.810     0.350     0.207  
 56602.450    45.660     0.330     0.767  
 56603.373   -11.750     0.330    -1.298  
 56603.449   -15.800     0.270     0.408  
 56604.386   -57.830     0.470    -0.087  
 56604.386   -57.880     0.430    -0.137  
 56604.527   -55.130     0.530     1.439  
 56606.435    46.210     0.270    -0.935  
 56606.450    46.590     0.400    -0.720  
 56606.450    46.510     0.340    -0.800  
 56607.453     7.850     0.310     0.007  
 56608.355   -50.350     0.270     0.820  
 56611.376    35.490     0.330     1.777  
 56612.274   -32.850     0.500    -3.376  
 56612.350   -34.790     0.280    -0.311  
 56613.343   -51.720     0.540     0.383  
 56613.398   -50.560     0.320    -0.504  

HIP 21443

 52214.654     8.060     1.370    -0.518  
 52215.635   -21.730     1.310     0.675  
 56223.530    -1.740     0.230     2.119  
 56224.634    -0.200     0.430     1.520  
 56226.559   -10.250     0.890     1.545  
 56231.586     8.970     0.300    -0.327  
 56235.598    15.050     0.300     0.131  
 56239.568    17.720     0.360    -0.988  
 56598.599    11.070     0.300    -0.378  
 56607.601   -31.010     0.350    -1.217  
 56612.607    15.200     0.370    -0.680  
 56613.423   -16.610     0.360    -1.276  

HIP 28678

 56223.572   -14.290     0.340     1.082  b
 56224.577     9.870     0.440    -0.512  a
 56224.577   -16.240     0.340    -0.702  b
 56231.598    12.150     0.320     0.742  a
 56231.598   -16.480     0.320     0.110  b
 56236.590    12.250     0.360     0.265  a
 56236.590   -17.680     0.400    -0.499  b
 56239.643    11.600     0.560    -0.641  a
 56239.643   -17.730     0.430    -0.287  b
 56241.570    11.120     0.720    -1.231  a
 56241.570   -17.090     0.470     0.466  b
 56602.630    27.710     0.370    -0.317  b
 56602.646   -31.910     0.450     0.229  a
 56603.568    30.680     0.520     1.710  b
 56603.584   -32.550     0.570     0.501  a
 56607.544   -32.790     0.350    -0.004  a
 56607.558    28.730     0.320     0.028  b
 56608.502    27.370     0.320    -0.646  b
 56608.517   -31.900     0.320     0.197  a
 56611.570    25.280     0.340     0.113  b
 56611.579   -29.730     0.330    -0.410  a
 56612.560    24.030     0.350    -0.133  b
 56612.570   -28.670     0.360    -0.331  a

HIP 41214

 56222.636    -4.100     1.390    -1.580  a
 56223.621    -4.860     1.270     0.687  a
 56236.631   -36.860     1.500     0.114  a
 56236.631   -16.260     2.880     0.016  b
 56241.621    -7.220     1.390    -0.652  a
 56241.621   -56.430     2.990    -0.561  b
 56604.610    -5.790     1.410    -0.084  a
 56604.623    -6.150     1.400    -0.483  a
 56604.623   -57.310     2.400    -0.267  b
 56606.634     0.600     1.260     1.538  a
 56606.640   -61.680     2.320     1.534  b
 56607.622   -65.720     1.430    -0.680  b
 56607.629     0.160     2.370    -0.322  a
 56608.612     0.840     1.280    -0.500  a
 56608.624   -65.820     2.450     0.355  b
 56611.637     0.250     1.360    -0.247  a
 56611.642   -64.080     2.560     0.981  b
 56611.648   -65.550     2.510    -0.499  b
 56612.653    -0.160     1.380     0.987  a
 56612.660   -62.870     2.600     0.040  b

HIP 96434

 56222.225   -29.810    10.240     8.063  a
 56223.171   -26.170     2.300    -3.293  b
 56224.229   -42.390     2.860    -0.490  a
 56224.229   -20.530     2.560     0.263  b
 56226.192   -44.230     2.460     0.519  a
 56226.192   -18.070     2.320    -0.294  b
 56228.213   -46.450     2.440     0.285  a
 56228.213   -16.540     2.360    -0.867  b
 56232.201   -49.750     2.460    -1.744  a
 56232.201   -14.740     2.290    -0.413  b
 56235.187   -46.340     2.550    -0.177  a
 56235.187   -16.230     2.290     0.048  b
 56240.172   -32.070    10.320     0.477  a
 56526.470   -43.160     0.870    -0.656  a
 56526.470   -19.760     0.640     0.393  b
 56532.425   -46.950     0.600     0.971  a
 56532.425   -14.780     0.290    -0.364  b
 56534.294   -47.310     0.550     0.655  a
 56534.294   -13.500     0.330     0.870  b
 56537.366   -45.470     0.600     0.323  a
 56537.366   -16.910     0.310    -0.240  b
 56538.325   -45.900     0.830    -1.567  a
 56538.325   -18.510     0.440    -0.294  b
 56540.305   -26.370    10.820    -3.095  b
 56540.321   -29.340    10.450    10.165  a
 56542.326   -29.740    10.230     1.376  a
 56543.309   -33.260    10.300    -8.010  a
 56544.264   -17.600     0.490     0.812  a
 56544.264   -46.150     0.270    -0.483  b
 56545.315   -10.860     0.320    -0.690  a
 56545.329   -10.020     0.460     0.042  a
 56545.329   -54.010     0.270     0.500  b
 56546.405   -62.630     0.480    -0.239  b
 56546.418     0.270     1.130     2.816  a
 56598.194   -13.470     4.570     1.498  b
 56598.210   -46.310     2.290     1.099  a
 56598.210   -14.530     4.230     0.429  b
 56600.172   -12.980     2.650     1.315  b
 56600.189   -47.970     2.390     0.067  a
 56600.189   -14.760     2.540    -0.466  b
 56601.182   -45.240     4.030     2.762  a
 56601.182   -16.850     2.420    -2.520  b
 56602.194   -12.480     2.400     2.176  b
 56602.207   -46.110     2.480     1.579  a
 56602.207   -13.270     2.250     1.393  b
 56602.223   -50.240     2.760    -2.559  a
 56602.223   -14.800     2.380    -0.130  b
 56603.196   -46.460     2.490     0.620  a
 56603.196   -14.010     2.280     1.297  b
 56604.221   -16.450     2.380    -0.080  b
 56604.233   -47.780     2.450    -1.718  a
 56604.233   -16.520     2.380    -0.135  b
 56605.304   -15.860     2.350     2.163  b
 56605.311   -45.210     2.420    -0.706  a
 56605.311   -17.510     2.410     0.525  b
 56606.247   -18.840     2.300     1.169  b
 56606.254   -41.880     2.760     0.744  a
 56606.254   -19.730     2.540     0.296  b
 56606.261   -43.620     2.500    -1.012  a
 56606.261   -19.300     2.430     0.743  b
 56607.171   -46.250     2.450    -6.027  a
 56607.171   -26.660     2.320    -4.091  b
 56607.183   -41.080     2.310    -0.893  a
 56607.183   -23.850     2.220    -1.243  b
 56608.178   -44.670     2.460    -7.894  a
 56608.178   -29.460     2.310    -3.241  b
 56611.164   -17.220     2.910     2.581  a
 56611.164   -43.560     2.420     0.635  b
 56612.176   -51.420     2.310     1.127  b
 56612.186   -11.400     2.300     0.437  a
 56613.176   -59.760     2.300     0.612  b
 56613.190    -5.150     2.340    -0.715  a

\end{verbatim}

}

\bsp

\label{lastpage}

\end{document}